# H2O: An Autonomic, Resource-Aware Distributed Database System

Angus Macdonald, Alan Dearle, Graham NC Kirby

{angus,al,graham}@cs.st-andrews.ac.uk

Jack Cole Building, School of Computer Science,
University of St Andrews, Fife, KY16 9SX

**Abstract.** This paper presents the design of an autonomic, resource-aware distributed database which enables data to be backed up and shared without complex manual administration. The database, H2O, is designed to make use of unused resources on workstation machines.

Creating and maintaining highly-available, replicated database systems can be difficult for untrained users, and costly for IT departments. H2O reduces the need for manual administration by autonomically replicating data and load-balancing across machines in an enterprise.

Provisioning hardware to run a database system can be unnecessarily costly as most organizations already possess large quantities of idle resources in workstation machines. H2O is designed to utilize this unused capacity by using resource availability information to place data and plan queries over workstation machines that are already being used for other tasks.

This paper discusses the requirements for such a system and presents the design and implementation of H2O.

## 1 Introduction

Providing replication and load-balancing in modern database systems can be difficult for untrained users, while the creation of database applications incurs substantial setup and maintenance costs if they are required to scale and be resilient to failure. Servers running such database systems cost a substantial amount to support [1], and may be unnecessary since most organizations have large amounts of unused resources already available on workstation machines [2].

We propose to make better use of local resources by creating a database which uses the existing infrastructure of an organization to provide database services. Our system automates the replication and load-balancing of small-scale databases to reduce the need for database administration, enables the sharing of data, and exploits existing resources on unused machines to maximize use of local resources and minimize reliance on expensive server room clusters.

The remainder of this paper is organized as follows. Section 2 discusses the motivation for our work. Section 3 derives requirements from these motivations which are used to guide the design of the system in Section 4. Section 5 discusses our current implementation of H2O, while Section 6 discusses work that is still to be done. Finally Sections 7 and 8 present related work and our conclusions.

## 2 Motivation

We are motivated by a number of observations on the use of database systems and workstations within enterprises:

1. *Sharing*. Allowing data to be shared is difficult without a central solution.
2. *Availability*. Ensuring data is always available forces a user to implement a replication strategy which tolerates machine failure.
3. *Administration*. Replication strategies for databases require significant manual administration, much of which could be avoided.
4. *Utilization*. Workstation machines are often underused and could be utilized to run distributed applications.

Many databases start small on a single user's machine. For example, in a University department a secretary may create and maintain a small database of student queries, while an admissions officer may create another database of applicants. The users of these databases may want to do two things: first, ensure the database is backed up to guard against machine failure, and second, to share the data with another member of staff.

The University may have a backup plan which ensures database files are replicated, but this only guarantees a recent copy of data, not necessarily the current copy. Such backups don't address the need for data to be shared. The user could provide remote access to the database on their machine, but as more people become involved this becomes undesirable, and if the machine is turned off the data becomes unavailable. The user could request that their IT department provides a centralized database solution, but this may be time consuming to set up and require frequent maintenance. Data would also have to be migrated to the new database before it could be used.

The proposed system solves the backup problem by providing database services which automatically replicate data onto machines in the same enterprise. Data can be easily shared with other members of staff, who can access it transparently without regard for its location. Replication ensures that the data remains available even when the original user takes their machine offline, and makes it more resilient against hardware failure.

The system will reduce the need for manual administration by automatically creating replicas for resilience and availability. If many users are accessing the same data then the system will autonomically balance the load between these replicas and adjust the location and replication factor of the data itself.

By providing a method to harness otherwise unused resources the system increases the overall computational and storage capacity of organizations which run it.

## 3 Requirements

A database system designed to run over existing infrastructure must meet the following requirements with respect to our initial motivations:

1. *Sharing: access to data must be location-independent*. Users should be able to share and access data without regard for its location.
2. *Availability: the system must be highly available.* Data should be replicated to limit unavailability as a result of the failure or periodic unavailability of individual machines. Data should be replicated sufficiently to be made resilient to permanent failure (e.g. disk loss).
3. *Administration: the system must be self-managing.* Changes in the availability of machines, the availability of resources, and access patterns require that data can be replicated and moved dynamically. The database system must be able to automatically make these adjustments without manual administration.
4. *Utilization: the system must be resource-aware.* To utilize unused resources effectively on workstations the system must be able to identify what resources are available. This requires a resource monitoring framework to collect and collate monitoring data.
5. *Compatibility: the system must be comparable to centralized databases.* A centralized database system will typically aim to provide fast response times to queries but may require substantial effort to scale up and manage replication. The proposed system should provide the same ACID transactions as a centralized solution, with comparable query response times. It should show a perceptible improvement in the time taken to manage other factors involving maintenance of the data itself.

## 4 Design

Each of the requirements outlined above guides the design of the database system:

1. *Access to data must be location-independent*. Users interact with the database system through an interface on their local machine. A location-aware database substrate is responsible for locating and querying data.
2. *The system must be available.* Tables are replicated over many machines so that the failure or unavailability of one machine does not render the database unavailable.
3. *The system must be self-managing.* Resource monitoring data is used in query planning and data placement decisions. Autonomic components monitor various aspects of the system's state and make changes to data placement and replication factors when necessary.

4. *The system must be resource-aware.* Each machine hosts a resource monitor to capture resource availability. Another monitoring component is able to collate this monitoring data to use it in query planning and data placement.
5. *The system must be comparable to centralized databases.* Users access the database system through a JDBC interface on their own machine. Two-phase locking of replicas is used to ensure strict serializability as part of the database's support of ACID transactions.

### 4.1 Architecture

The database system consists of a database instance running on every available machine in an enterprise. Relations can be replicated onto multiple machines, with the copies of data for each relation managed by a Table Manager, which is responsible for the locking and persistence of the table it is managing. To enable discovery of existing tables and to mediate the creation of new tables, the system maintains a System Table which holds references to all extant Table Managers. Users submit queries to the database instance on their machine and the query is executed at the most appropriate replicas. This architecture is illustrated below in Figure 1.

In this example machine B is responsible for maintaining the System Table which holds references to the Table Managers for *X* and *Y*. The Table Manager for table *X* maintains references to replicas of the table data on *C* and *D*, while the Table Manager for *Y* keeps references to the data on *C*. A user making queries from machine A has no knowledge of the location of the System Table, the Table Managers, or the data.

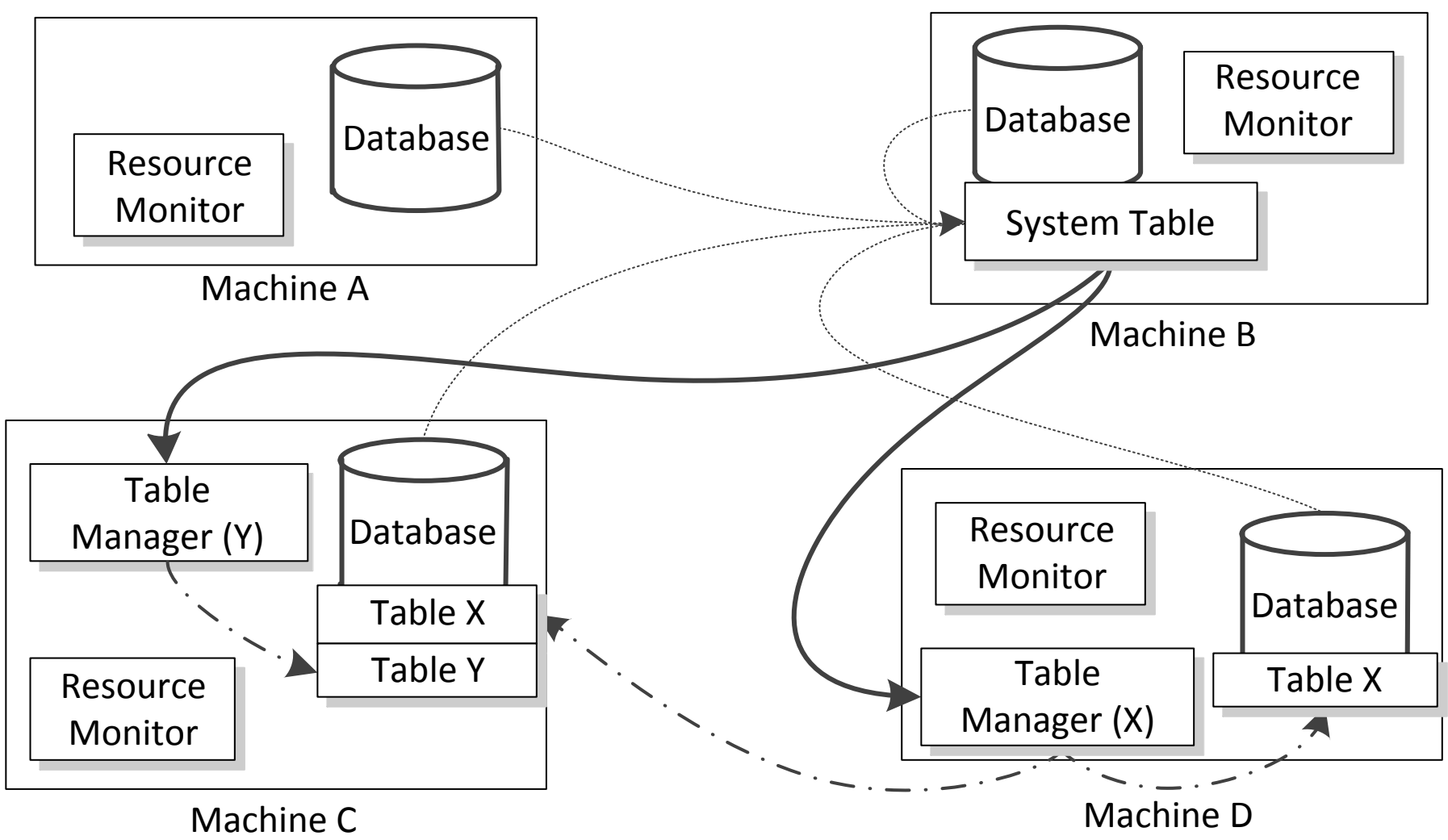

**Fig. 1.** Overview of System Architecture.

The System Table is needed to find extant relations, though references to these relations may also be cached locally by database instances. The System Table is effectively a write-through cache whose state is synchronously replicated following changes, so that it can fail without affecting the availability of the database system. Similarly, Table Managers are replicated synchronously, though locking information is not persisted because in the event of failure any running transactions are rolled back. Because critical meta-data is replicated, the system is able to recover from failure by re-instantiating the System Table and Table Managers on other available machines.

The strict two-phase locking approach used in H2O ensures serializability. To ensure replicas are kept consistent two-phase commit is used on updates.

### 4.2 Example Query

To illustrate the architecture of the system consider how a basic join query is executed by the database system.

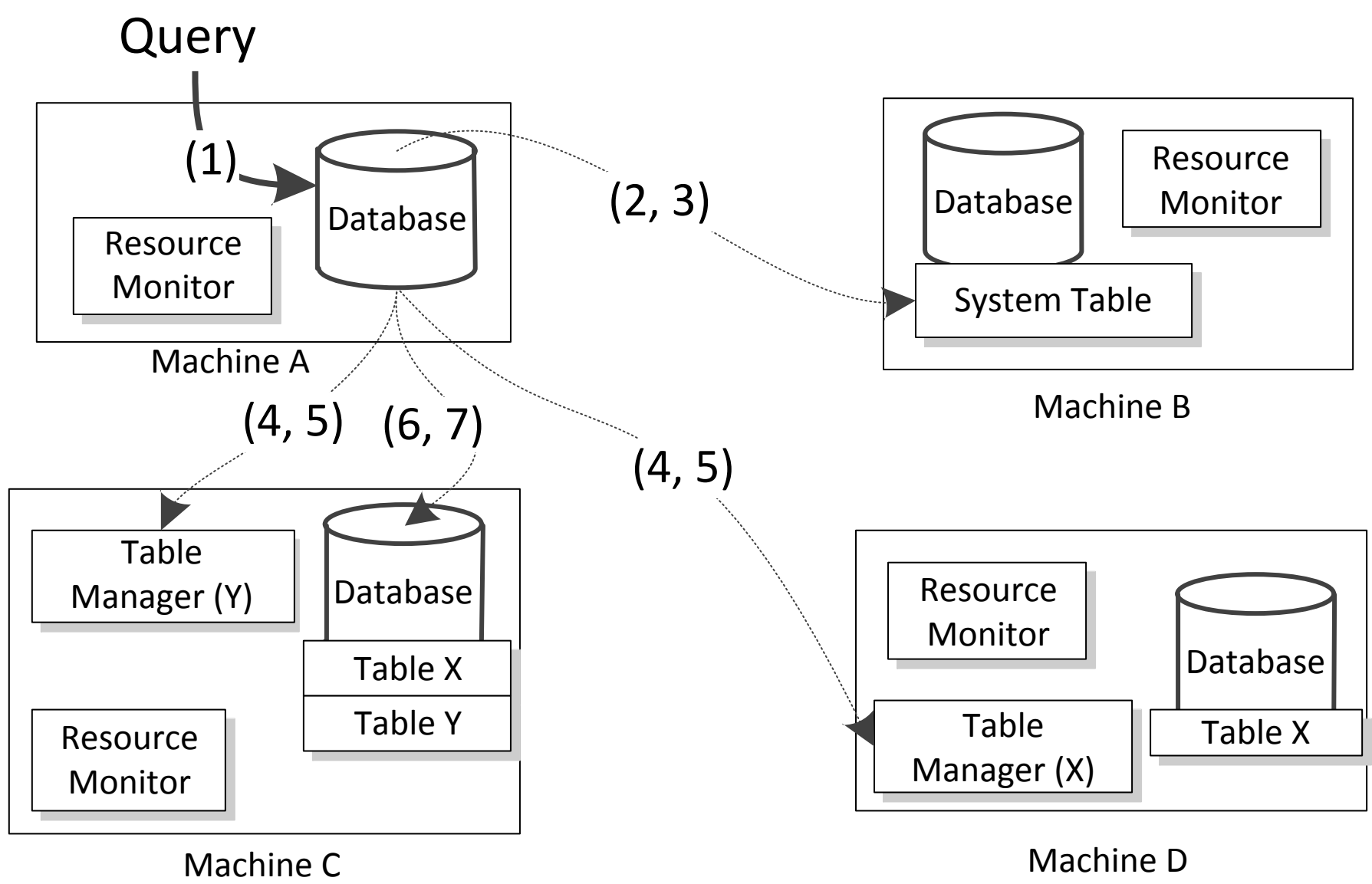

**Fig. 2.** Example Join Query.

1. A user submits a query via a database interface on their machine, *A*.

   `SELECT * FROM X, Y WHERE X.a_id = Y.a_id;`

2. Their local database instance (on machine A) parses the query and sends a request to the System Table for the location of the *X* and *Y* Table Managers.
3. The System Table returns the location of these Table Managers on machines *C* and *D*.

4. The user's local database instance (which now has references to both Table Managers) requests read locks on both tables from their managers.
5. The Table Managers return locks and meta-data describing where the table data can be found.
6. The query is sent to machine *C*, which holds both tables, and is then executed. The decision about which machine executes the query is based on monitoring information relating to computational availability on machines and on database monitoring of aspects such as table size.
7. Once the query has been executed, read locks for both tables are released and the result of the query is returned to the user.

# 5 Implementation

H2O is an implementation of the resource-aware distributed database system described above. It is designed to run over small sets of workstation machines (tens to low hundreds). The main database functionality of H2O is provided by the H2 database system [3] around which the rest of the system is built.

Each database instance consists of an H2 database modified to support replication, and a resource monitor. One of these instances manages the System Table and each instance may have many extant Table Manager processes.

## 5.1 Bootstrapping

The first database instance to be started is responsible for creating the System Table. Subsequent instances connect by specifying the location of a known instance, which is used as an entry point into the system. We use an implementation of the peer-to-peer overlay Chord [4] to provide this bootstrapping mechanism. Chord's lookup functionality is used to find the System Table on startup, thus abstracting over locality.

## 5.2 Detecting Failure

Chord's maintenance mechanism is used as a means of detecting database instance failure. On the failure of a database instance the succeeding instance in the Chord ring detects this failure and makes an up-call into H2O. The database notifies Table Managers that may have had replicas located on the failed instance, allowing them to create more replicas if necessary.

When the System Table or a Table Manager has also failed it is re-instantiated elsewhere from persisted copies of its state on the successor node.

## 6 Future Work

The architecture described above is currently implemented. This section presents the remainder of our design as a series of open-ended design decisions.

Resource monitoring systems typically run processes on each machine to collect raw data, and use logically centralized databases to collate data for processing [5]. H2O will use this approach, storing monitoring data in the database itself.

Every instance will monitor CPU, memory and disk utilization, while every Table Manager will monitor replication factor and access patterns. These access patterns include factors such as query response time, load balancing between replicas, and more general query patterns such as read-write ratios and burstiness.

The database system must be able to cope with change relating to the failure and variable resource availability of machines, and to usage patterns and demand. Changes will be made autonomically [6] by processes making use of monitoring data and tunable heuristics. Where possible, these processes should be decentralized to avoid the system becoming reliant on any one machine collecting all meta-data. For instance, replication factor can be trivially decentralized by making Table Managers responsible for them. Other decisions, such as those involving data placement, will be made by a system-wide autonomic process when more global information is needed.

Due to the potential quantity of resources made available on workstation machines, many may remain unused even in the presence of a resource-aware database system. These resources can be speculatively harnessed to test placement strategies and other non-critical operations. Data can be replicated onto unused machines to be reformatted and repartitioned speculatively, then queries can then be executed against both the primary and speculative replicas as a means of evaluating possible placement strategies and updating the system's knowledge-base.

Failure of databases instances can take one of a number of forms. It can be either unexpected, where the process dies, or pre-empted, where the process anticipates that there will soon be too few resources available to service requests. When failure is pre-empted processes can be migrated to more available machines, as is possible with Condor [7]. Unexpected failure can be permanent, meaning data is unrecoverable (for example, as a result of disk loss), or transitory, meaning data will become available at an undetermined later point (for example on machine restart, or after a power cut).

## 7 Related Work

Condor [7] is a scheduling system which aims to maximize the utilization of workstation resources by allowing long running computations to be run remotely. Processes are checkpointed so that they can be paused for a short period, or moved between machines (process migration). Compute jobs are self-contained, and so can be restarted on any available workstation. Data is maintained on the machine which submitted the compute job and not on the machine running the job, to prevent the job from monopolising resources on the remote machine.

H2O differs from Condor in that it is explicitly aiming to store relatively large quantities of data on workstation machines. It is the data and not the computation which is the focus of our work. Brief periods of inactivity are acceptable in the context of long-running computations, but not in database systems where quick responses are demanded.

Research looking at server power consumption has shown that machines which are not used to full capacity consume a substantial portion of their peak power consumption [8]. For instance, servers with near zero percent utilization still use around 50% of the power used at peak utilization. A consequence of this consumption is that resources left unused are a considerable source of waste. H2O aims to address this by making better use of existing resources.

There is a substantial quantity of work on data placement and query optimization in database systems generally [9][10]. While centralized decision making is most typical, various approaches have been taken to spread decision making through the system. Mariposa [11] uses a micro-economic approach to decentralizing decision-making in a heterogeneous wide-area DDBMS. Nodes operate in a market economy, buying and selling space for replicas, and time for queries. Consequently, they are able to indicate their resource availability through pricing. H2O is designed to run over local area networks, so it doesn't require decision making to be decentralized to the same degree as Mariposa.

Piazza [12] introduced 'spheres of co-operation' designed to cluster heterogeneous databases together to make decisions about query optimization as a group. Grouping is seen as more scalable than global optimization, but provides a broader knowledge-base than purely local optimization. This solution may be appropriate for H2O if system-wide decision making proves impractical or consumes too many local resources.

Commercial database systems use a variety of mechanisms to support replication. Oracle [13] supports snapshot replication from master databases, where snapshots may be a subset of the database. If there are multiple masters updates can be propagated synchronously as they occur, or asynchronously via batch update. Microsoft SQL Server [14] uses a publish-subscribe model for replication where subscription can be either push or pull. Snapshot replication is supported as well as transactional replication, where individual updates are propagated, and merge replication, where databases act autonomously and are later merged. H2O provides synchronous transactional replication managed through Table Managers (primary copy locking [9]), though the system's design is not dependent on this mechanism.

More recent database systems such as Greenplum [15] and Aster [16] are based on shared-nothing architectures that allow new nodes to be added with a linear increase in performance. Both systems automatically partition data across nodes, attempting to minimize intra-node data transfer [17]. They are designed to be run on server clusters, not workstation machines, though the mechanisms used to add nodes and to partition data are relevant to the design of H2O.

There are numerous resource monitoring systems aimed at grid computing [5] which provide the functionality needed for this project. H2O is more notable for its use of resource monitoring data in database decision making. Clustered database systems generally assume machine-level resources are static and so do not factor dynamic resource availability into query planning. More recently some projects have begun looking at energy consumption as a primary performance metric in query planning [18]. Mariposa indirectly factors resource utilization into query planning through pricing.

While clustered database systems tend to support ACID semantics, wider-area databases tend to sacrifice the consistency for reasons captured by the CAP theorem [19]. If consistency is to be prioritized, availability or partition tolerance must be sacrificed. In a clustered database system where partitions are rare, availability is favoured over partition tolerance. Conversely, systems that operate over a wide area tend to sacrifice consistency because of the likelihood of partitions and machine failure.

PIER [20], which supports hundreds of thousands of machines, offers no guarantees about the freshness of data received as part of a query. Amazon SimpleDB [21], one of many cloud offerings, provides eventual consistency with an option for consistent reads. H2O aims to provide ACID transactions by operating at a smaller scale over local area networks.

Yang et al. [22] present a database clustering middleware which runs clusters of databases running on off-the-shelf hardware. Each machine runs a single MySQL database, which is expected to be large enough to support an entire web application. Updates are made to a number of replicas using the two-phase commit protocol, meaning the system can recover from individual machine failure. ACID-compliant transactions are also supported. This shows that centralized databases can scale out to support synchronous replication and ACID-compliant transactions. H2O differs in running over workstation machines with variable resources.

# 8 Conclusion

This paper has presented the requirements for a resource-aware database system, and outlined the design and implementation of H2O which aims to meet these requirements. H2O is currently an operational database system and is able to replicate data across many machines with manual administration. We are currently developing the autonomic functionality of the database in order to automate this decision making.

Once development is complete, we will begin evaluations of the system's performance. These evaluations will look at the performance of the database when answering queries, the effectiveness of the resource monitor at detecting availability and the autonomic system in adapting to change.

At a point where resource consumption is a growing concern H2O is notable in using existing resources to accomplish a task normally achieved with clusters of computers in server rooms. This may help to reduce energy consumption within organizations.

This work aims to show that it is possible for a database to be distributed over a set of workstations without need for manual administration. We believe that this approach of utilizing workstations is suitable for a wider class of applications, provided they can be adapted to take resource availability into account in decision making.

## 9 References

[1] A. Greenberg, J. Hamilton, D. Maltz, and P. Patel, "The cost of a cloud: research problems in data center networks," ACM SIGCOMM Computer Communication Review, vol. 39, 2008, p. 68–73.

[2] M. Mutka and M. Livny, "Profiling workstations' available capacity for remote execution," Performance, 1987, p. 529–544.

[3] M. Thomas, "H2 Database." (http://www.h2database.com)

[4] I. Stoica, R. Morris, D. Karger, M. Kaashoek, and H. Balakrishnan, "Chord: A scalable peer-to-peer lookup service for internet applications," Proceedings of the 2001 conference on Applications, technologies, architectures, and protocols for computer communications, ACM, 2001, p. 160.

[5] S. Zanikolas and R. Sakellariou, "A taxonomy of grid monitoring systems," Future Generation Computer Systems, vol. 21, 2005, p. 163188.

[6] J. Kephart and D. Chess, "The vision of autonomic computing," Computer, vol. 36, 2003, pp. 41-50.

[7] M. Litzkow, M. Livny, and M. Mutka, "Condor-a hunter of idle workstations," proceedings of the 8th International Conference of Distributed Computing Systems, 1988.

[8] L.A. Barroso and U. Hölzle, "The Case for Energy-Proportional Computing," Computer, vol. 40, 2007, pp. 33-37.

[9] M. Özsu and P. Valduriez, Principles of Distributed Database Systems, 1999.

[10] J. Hellerstein, "Architecture of a Database System," Foundations and Trends in Databases, vol. 1, 2007, pp. 141-259.

[11] M. Stonebraker, P.M. Aoki, W. Litwin, A. Pfeffer, A. Sah, J. Sidell, C. Staelin, and A. Yu, "Mariposa: a wide-area distributed database system," The VLDB Journal The International Journal on Very Large Data Bases, vol. 5, 1996, pp. 48-63.

[12] S. Gribble, A. Halevy, Z. Ives, M. Rodrig, and D. Suciu, "What can databases do for peer-to-peer," WebDB Workshop on Databases and the Web, Citeseer, 2001.

[13] Oracle, "Oracle Database 11g."

[14] Microsoft, "Microsoft SQL Server." (http://www.microsoft.com/sqlserver/2008/en/us/)

[15] Greenplum, Greenplum Database 4.0, 2009. (http://www.greenplum.com)

[16] T. Argyros, M. Bawa, and G. Candea, Next-Generation Data Warehouses, 2008.

[17] D.J. DeWitt and J. Gray, "Parallel database systems," ACM SIGMOD Record, vol. 19, 1990, pp. 104-112.

[18] W. Lang and J. Patel, "Towards eco-friendly database management systems," Imprint, 2009.

[19] S. Gilbert and N. Lynch, "Brewer's conjecture and the feasibility of consistent, available, partition-tolerant web services," ACM SIGACT News, vol. 33, 2002, p. 51.

[20] R. Huebsch, B. Chun, J. Hellerstein, B. Loo, P. Maniatis, T. Roscoe, S. Shenker, I. Stoica, and A. Yumerefendi, "The architecture of PIER: an internet-scale query processor," Proceedings of 2nd Conference on Innovative Data Systems Research (CIDR), 2005, p. 2843.

[21] Amazon.com, "Amazon SimpleDB," http://aws.amazon.com/simpledb/.

[22] F. Yang and R. Yerneni, "A Scalable Data Platform for a Large Number of Small Applications," research.yahoo.net, 2009.